\documentstyle[12pt]{article}
\def\calr{ {\cal R} }
\def\cl{ {\cal L} }   
\def\ct{ {\cal T}}
\begin{document}
\vspace{2cm}
\title{Analytic Bethe Ansatz for 1-D Hubbard model and
twisted coupled XY model }
\author{ Ruihong Yue and Tetsuo Deguchi
	  \\[.3cm]
      Department of Physics  \\
      Faculty of Science\\
      Ochanomizu University \\
      112 Tokyo, Japan}

\date{\today}
\maketitle
\begin{abstract}
We found the eigenvalues of the transfer matrices for the 
1-D Hubbard model and for the coupled XY model with twisted 
boundary condition by using the analytic Bethe Ansatz method. 
Under a particular condition the two models  have the same 
Bethe Ansatz equations. We have also proved that the periodic 
1-D Hubbard model is exactly equal to the coupled XY model 
with nontrivial twisted boundary condition at the level of
hamiltonians and transfer matrices. 

\vspace{.5cm}\noindent
\end {abstract}
\raisebox{15cm}[][]{\hspace{12cm} \bf  OCHA-PP-81\\}
\newpage
\bigskip
\section{Introduction}

The 1-D Hubbard model (the Hubbard model) is one of the 
significant exactly solvable models in condensed matter 
physics. Lieb and Wu \cite{LW} succeeded in  diagonalizing 
the hamiltonian in the frame of the coordinate Bethe Ansatz. 
However, the integrability of the Hubbard model was recently 
set up by Shastry \cite{Sha1,Sha2}, Olmedilla, Wadati and 
Akutsu \cite{WOA,OWA,OW} from the viewpoint of the  Quantum 
Inverse Scattering Method (QISM). The Yang-Baxter equation 
for the related R matrix was proved in \cite{SW}. In
\cite{Sha1,Sha2}, Shastry found the Yang-Baxter relation and
constructed the transfer matrix of the coupled XY model which
is equal to the Hubbard model with the  help of the 
Jordan-Wigner transformation. The  commutative family with 
one free parameter ensures the integrability of the system. 
In a different approach, Olmedilla, Wadati and Akutsu 
\cite{WOA,OWA,OW},  starting from the super L-operator of 
the Hubbard  model, solved the super-Yang-Baxter (SYB) relation 
and found  the invertible R matrix which is the same as that 
given by Shastry \cite{Sha3} up to a scalar function. In both 
cases \cite{Sha2,OWA}, the hamiltonian can be derived
from the transfer matrix under the periodic boundary condition.

 The base of the Quantum Inverse Scattering Method  is the
Yang-Baxter relation. The later has closely  related to the 
Quantum group and the Yangian. This technique provides a 
systematic method to deal with the integrable 1-D quantum 
systems and 2-D solvable statistical mechanical models. It is 
known that most of integrable systems can be handled in the 
frame of the   Algebraic Bethe Ansatz ( or Analytic Bethe 
Ansatz ) method. But, up to now, there is no report about 
the diagonalization of the transfer matrix for the 
Hubbard model in the QISM approach. So, it is important to 
find the  solution of the Hubbard model by using QISM. 
In \cite{Sha3}, Shastry conjectured the eigenvalue
of the transfer matrix of the coupled XY model based upon 
the coordinate Bethe Ansatz method.  Comparing the Bethe 
Ansatz equations given by Lieb and  Wu \cite{LW} and by 
Shastry \cite{Sha3}, an  extra factor appears and it shows 
some difference between the two models.

  The motivation of this paper is to find the eigenvalues 
of the transfer matrices  related to the Hubbard model and 
the coupled XY model with twisted boundary condition by a 
version of analytic Bethe Ansatz method (ABA). We want 
to give some explanation about the difference between the 
coupled XY and the Hubbard models. The extra factor 
originates  from the boundary condition. In fact the 
Jordan-Wigner transformation does not keep the invariance 
of the boundary condition.  We also show that the Hubbard 
model with periodic boundary condition is exactly equal to 
the coupled XY model with a special boundary condition.

The organization of this paper is to recall the Yang-Baxter 
relation for the Hubbard model  and the parametrization of the
R matrix in section 2. In section 3, we  investigate the model
at special cases by using the algebraic Bethe Ansatz method.
Because there are two kinds of creation operators with spin-up
( or spin-down ), such as $T_{21}$ and $T_{43}$  ($T_{21}$ and 
$T_{43}$),  the general  multi-particle states become very 
complicated. However, the special solution gives an insight  
about the general structure of the eigenvalue. Section 4 will 
attributes  to an  analytic Bethe Ansatz method to the  Hubbard 
model. In our approach, the analytic property of the eigenvalue  
of the transfer matrix, together with the asymptotic behaviour,  
determines almost all   unknown functions. The discussion 
is different from the standard  analytic Bethe  Ansatz method 
since there is not the  crossing symmetry,   which has played 
an important role \cite{Res1,VR,Res2,Mar}. In section 5, we 
will apply  the analytic Bethe Ansatz method developed in 
section 4 to the coupled XY model with twisted boundary 
condition.   Under a special choice, it recovers the results 
of the  Hubbard  model in the level of Bethe Ansatz equations 
and the eigenvalue. In section 6 we show  that the Hubbard 
model with periodic boundary condition is equal to the  coupled 
XY model  with special boundary condition by using the 
Jordan-Wigner   transformation and some discussions are given 
in section 7.

  \section{ The Hubbard model and the Super-Yang-Baxter relation}

In this section we  recall the definition of the SYB   relation 
for the Hubbard model and some useful functional  relations. 
We follow the  notation  in \cite{OWA}. The  hamiltonian of the 
Hubbard model is 
\begin{equation}
H_{Hu}=-\sum_{j=1,s=\uparrow,\downarrow}^L
        \left(a^+_{m+1,s}a_{m,s}+a^+_{m,s}a_{m+1,s}\right)
	  +U\sum_{m=1}^L(n_{m\uparrow}-1/2)(n_{m\downarrow}-1/2)
	  \end{equation}
where $a^+_{m,s}$ $(a_{m,s})$ stands for the $m$-th site electron
creation ( annihilation ) operator with spin $s$. The 
Super-Yang-Baxter relation is \cite{OWA}
\begin{equation}
\calr(\mu,\nu)\left[\cl_m(\mu)\otimes_s \cl_m(\nu)\right]
=\left[\cl(\nu)_m\otimes_s\cl_m(\mu)\right]\calr(\mu,\nu)
\end{equation}
The super tensor product is defined by
\begin{equation}
\left[A\otimes_s B\right]^{ia}_{jb}=
 A^i_jB^a_b(-1)^{[p(i)+p(j)]p(a)},
\end{equation}
where $p(1)=p(4)=-p(2)=-p(3)=1$ are the parities. 
For the Hubbard model the L-operator takes the form
\begin{eqnarray}
\lefteqn{\cl(\mu)=}  \nonumber\\ & &
\left(\begin{array}{cccc}
  -e^{h(\mu)}f_{m\uparrow}f_{m\downarrow} &
  f_{m\uparrow}a_{\downarrow} &
  ia_{m\uparrow}f_{m\downarrow} &
  ia_{m\uparrow}a_{m\downarrow}e^{h(\mu)}\\[3mm]
  -if_{m\uparrow}a^+_{m\downarrow}&
 f_{m\uparrow}g_{m\downarrow}e^{-h(\mu)}&
 a_{m\uparrow}a^+_{m\downarrow}e^{-h(\mu)}& 
ia_{m\uparrow}g_{m\downarrow}\\[3mm]
a^+_{m\uparrow}f_{m\downarrow} &
a^+_{m\uparrow}a_{m\downarrow}e^{-h(\mu)}&
g_{m\uparrow}f_{m\downarrow}e^{-h(\mu)}&
g_{m\uparrow}a_{m\downarrow}\\[3mm]
-ia^+_{m\uparrow}a^+_{m\downarrow}e^{h(\mu)}&
a^+_{m\uparrow}g_{m\downarrow}&
ig_{m\uparrow}a^+_{m\downarrow}&
-g_{m\uparrow}g_{m\downarrow}e^{h(\mu)}
  \end{array}\right)
  \end{eqnarray}
where $\mu$ is the spectrum parameter,  and the functions 
$f_{ms}$ and $g_{ms}$ are  
\begin{equation}
\begin{array}{c}
\begin{array}{rcl}
f_{ms}&=&w_4(\mu)-w_3(\mu)-\{w_4(\mu)-w_3(\mu)-
i[w_4(\mu))+w_3(\mu)]\}n_{ms} \\[3mm]
g_{ms}&=&w_4(\mu)+w_3(\mu)-\{w_4(\mu)+w_3(\mu)-
i[w_4(\mu)-w_3(\mu)]\}n_{ms}
\end{array}\\[3mm]
\begin{array}{rcl}
w_4(\mu)+w_3(\mu)&=&\sin (\mu+\pi/4)\\[3mm]
w_4(\mu)-w_3(\mu)&=&\sin (\mu-\pi/4)\\[3mm]
\sinh (2h(\mu))&=&\displaystyle -\frac{U}4\cos (2\mu)
\end{array}
\end{array}
\end{equation}
The graded $\calr$  matrix reads
\begin{equation}\left(
\begin{array}{cccccccccccccccc}
\rho_1&0&0&0    &0&0&0&0         &0&0&0&0          &0&0&0&0 \\
0&\rho_2&0&0    &i\rho_9&0&0&0   &0&0&0&0          &0&0&0&0 \\
0&0&\rho_2&0    &0&0&0&0         &i\rho_9&0&0&0    &0&0&0&0 \\
0&0&0&\rho_3    &0&0&\displaystyle \frac{\rho_6}i&0  &0&i\rho_6&0&0  
                              &\rho_8&0&0&0 \\
0&\displaystyle \frac{\rho_{10}}i&0&0  &\rho_2&0&0&0  &0&0&0&0    
                           &0&0&0&0 \\
0&0&0&0       &0&i\rho_4&0&0     &0&0&0&0          &0&0&0&0 \\
\displaystyle
0&0&0&i\rho_6  &0&0&\rho_5&0   &0&\rho_7&0&0  
                               &\frac{\rho_6}i&0&0&0 \\
0&0&0&0    &0&0&0&\rho_2   &0&0&0&0    &0&\displaystyle 
          \frac{\rho_{10}}i&0&0 \\
0&0&\displaystyle \frac{\rho_{10}}i&0  &0&0&0&0    &\rho_2&0&0&0     
&0&0&0&0  \\ 
0&0&0&\displaystyle \frac{\rho_6}i    &0&0&\rho_7&0  &0&\rho_5&0&0  
                                        &i\rho_6&0&0&0 \\
0&0&0&0         &0&0&0&0      &0&0&\rho_4&0     &0&0&0&0 \\
0&0&0&0         &0&0&0&0      &0&0&0&\rho_2     
                         &0&0&\displaystyle \frac{\rho_{10}}i&0 \\
0&0&0&\rho_8    &0&0&i\rho_6&0  &0&\displaystyle \frac{\rho_6}i&0&0  
                                       &\rho_3&0&0&0 \\
0&0&0&0    &0&0&0&i\rho_9   &0&0&0&0    &0&\rho_2&0&0 \\
0&0&0&0         &0&0&0&0      &0&0&0&i\rho_9     
                               &0&0&\rho_2&0 \\
0&0&0&0         &0&0&0&0      &0&0&0&0  &0&0&0&\rho_{1} 
\end{array} \right)
\end{equation}
where all functions are defined  by
\begin{equation}
\begin{array}{rcl}
\rho_1(\mu,\nu)&=&\displaystyle e^l\alpha(\mu)\alpha(\nu)+
	         e^{-l}\gamma(\mu)\gamma(\nu), \\[5mm]
\rho_4(\mu,\nu)&=&\displaystyle  e^l\gamma(\mu)\gamma(\nu)+
                   e^{-l}\alpha(\mu)\alpha(\nu) ,\\[5mm]
\rho_9(\mu,\nu)&=& \displaystyle -e^l\alpha(\mu)\gamma(\nu)+
                   e^{-l}\gamma(\mu)\alpha(\nu) ,\\[5mm]
\rho_{10}(\mu,\nu)&=& \displaystyle e^l\gamma(\mu)\alpha(\nu)-
                   e^{-l}\alpha(\mu)\gamma(\nu) ,\\[5mm]
\rho_3(\mu,\nu)&=&\displaystyle
                  \frac{e^l\alpha(\mu)\alpha(\nu)
                  -e^{-l}\gamma(\mu)\gamma(\nu)}
                  {\alpha^2(\mu)-\gamma^2(\nu)},\\[5mm]
\rho_5(\mu,\nu)&=&\displaystyle
                  \frac{-e^l\gamma(\mu)\gamma(\nu)
                  +e^{-l}\alpha(\mu)\alpha(\nu)}
                  {\alpha^2(\mu)-\gamma^2(\nu)},\\[5mm]
\rho_6(\mu,\nu)&=&\displaystyle
                  \frac{e^{-h}[e^l\alpha(\mu)\gamma(\mu)
                  -e^{-l}\alpha(\nu)\gamma(\nu)]}
                  {\alpha^2(\mu)-\gamma^2(\nu)},\\[5mm]
\rho_7(\mu,\nu)&=&\rho_1(\mu,\nu)-\rho_3(\mu,\nu),\\[5mm]
\rho_3(\mu,\nu)&=&\rho_4(\mu,\nu)-\rho_5(\mu,\nu),\\[5mm]
l&=&h(\mu)-h(\nu),\\[3mm]
h&=&h(\mu)+h(\nu).
\end{array} 
\end{equation}

Due to the Super-Yang-Baxter relation,  one can define the 
monodromy matrix 
\begin{equation}
\ct(\mu)=\cl_L(\mu)\cdots\cl_2(\mu)\cl_1(\mu)
\end{equation}
which still satisfies  the Super-Yang-Baxter relation
\begin{equation}
\calr(\mu,\nu)\left[\ct(\mu)\otimes_s \ct(\nu)\right]=
\left[ \ct(\nu)\otimes_s \ct(\mu)\right]\calr(\mu,\nu)
\end{equation}
The Super-Yang-Baxter relation leads to  the  existence of 
the commutative family of the transfer matrices
$t_H(\mu)=str \ct(\mu) $ with infinitely many different value
of $\mu$. So, the Hubbard model is integrable. The infinite 
number of conserved quantities can be derived from the 
transfer matrix $t_H(\mu)$. The derivation of
$\log [t_H(\mu)]$ at $\mu=\pi/4$ gives  the hamiltonian of 
the  Hubbard model with the periodic boundary condition.
Before ending this section, we list some useful  functional
relations which will be used in the following sections,
\begin{equation}
\begin{array}{rcl}
1&=&\rho_1\rho_4+\rho_9\rho_{10},\\[3mm]
2&=&\rho_1\rho_5+\rho_3\rho_{4},\\[3mm]
1&=&\rho_3\rho_5-(\rho_6)^2,\\[3mm]
\rho_{10}&=&\rho_6(e^h\alpha(\mu)\alpha(\nu)
            +e^{-h}\gamma(\mu)\gamma(\nu)), \\[3mm]
\rho_{9}&=&\rho_6(e^{-h}\alpha(\mu)\alpha(\nu)
            +e^{h}\gamma(\mu)\gamma(\nu)). 
\end{array}
\end{equation}

\section{An algebraic analysis on the eigenvalue}

In this section, we will discuss the solution of the Hubbard model
in some special cases by using the algebraic Bethe Ansatz method. 
For the general multiparticle states, it becomes very difficult even 
if there are two electrons with opposite spins.

Taking the special elements of the Super-Yang-Baxter 
relation, one
can obtain the following  useful relations
\begin{eqnarray}
\ct_{44}(\mu)\ct_{43}(\nu)&=&\displaystyle
   \frac{\calr^{44}_{44}(\nu,\mu)}{\calr^{43}_{34}(\nu,\mu)}
           \ct_{43}(\nu)\ct_{44}(\mu)\nonumber\\
& & \displaystyle -\frac{\calr^{34}_{34}(\mu,\nu)}
    {\calr^{43}_{34}(\nu,\mu)} \ct_{43}(\mu)\ct_{44}(\nu)
\end{eqnarray}

\begin{eqnarray}
\ct_{33}(\mu)\ct_{43}(\nu)&=&\displaystyle
   \frac{-\calr^{33}_{33}(\mu,\nu)}{\calr^{43}_{34}(\mu,\nu)}
           \ct_{43}(\nu)\ct_{33}(\mu)\nonumber\\
& &\displaystyle -\frac{\calr^{43}_{43}(\nu,\mu)}{\calr^{43}_{34}
           (\mu,\nu)}\ct_{43}(\mu)\ct_{33}(\nu)
\end{eqnarray}

\begin{equation}
\begin{array}{rcl}
\ct_{22}(\mu)\ct_{43}(\nu)&=&\displaystyle
   -\left(\frac{\calr^{32}_{23}(\mu,\nu)}{\calr^{42}_{24}(\mu,\nu)}
         -\frac{\calr^{41}_{23}(\mu,\nu)\calr^{32}_{14}(\mu,\nu)}
           {\calr^{42}_{24}(\nu,\mu)\calr^{41}_{14}(\mu,\nu)}
            \right)\ct_{43}(\nu)\ct_{22}(\mu)\\[5mm]
& &\displaystyle+
\left(\frac{\calr^{14}_{23}(\nu,\mu)}{\calr^{42}_{24}(\mu,\nu)}
         -\frac{\calr^{14}_{23}(\nu,\mu)\calr^{14}_{14}(\mu,\nu)}
               {\calr^{42}_{24}(\nu,\mu)\calr^{41}_{14}(\mu,\nu)}
            \right)\ct_{41}(\nu)\ct_{24}(\mu)\\[5mm]
& &\displaystyle
   -\left(\frac{\calr^{23}_{23}(\nu,\mu)}{\calr^{42}_{24}(\mu,\nu)}
         -\frac{\calr^{41}_{23}(\nu,\mu)\calr^{23}_{14}(\mu,\nu)}
           {\calr^{42}_{24}(\nu,\mu)\calr^{41}_{14}(\mu,\nu)}
            \right)\ct_{42}(\nu)\ct_{23}(\mu)\\[5mm]
& &\displaystyle-
   \frac{\calr^{41}_{23}(\mu,\nu)\calr^{42}_{42}(\mu,\nu)}
        {\calr^{42}_{24}(\mu,\nu)\calr^{41}_{14}(\mu,\nu)}           
        \ct_{41}(\mu)\ct_{24}(\nu)\\[5mm] 
& &\displaystyle+
   \frac{\calr^{41}_{23}(\mu,\nu)}{\calr^{41}_{14}(\mu,\nu)}
           \ct_{21}(\mu)\ct_{44}(\nu)\\[5mm]   
& &\displaystyle+
   \frac{\calr^{42}_{42}(\mu,\nu)}{\calr^{42}_{24}(\mu,\nu)}
           \ct_{42}(\mu)\ct_{23}(\nu)
\end{array}
\end{equation}
\begin{equation}
\begin{array}{rcl}
T_{11}(\mu)T_{43}(\nu)&=&\displaystyle
   \frac{\calr^{31}_{13}(\mu,\nu)}{\calr^{41}_{14}(\mu,\nu)}
   \ct_{43}(\nu)\ct_{11}(\mu)+
   \frac{\calr^{13}_{13}(\mu,\nu)}{\calr^{41}_{14}(\mu,\nu)}
      \ct_{41}(\nu)\ct_{13}(\mu)\\[5mm]
& &\displaystyle
    +\frac{\calr^{41}_{23}(\mu,\nu)}{\calr^{41}_{14}(\mu,\nu)}
    \ct_{21}(\mu)\ct_{33}(\nu)
    + \frac{\calr^{41}_{32}(\mu,\nu)}{\calr^{41}_{41}(\mu,\nu)}
   \ct_{31}(\mu)\ct_{23}(\nu)\\[5mm] 
& &\displaystyle-
   \frac{\calr^{41}_{41}(\mu,\nu)}{\calr^{41}_{14}(\mu,\nu)}
           \ct_{41}(\mu)\ct_{13}(\nu)   
\end{array}
\end{equation}
   
For the Hubbard model the Hilbert space consists of four 
states: double occupied state $|\uparrow\downarrow\rangle$, 
spin-up state $|\uparrow\rangle$, spin-down state 
$|\downarrow \rangle$ and unoccupied state $|0\rangle$. We  
denote them by $|1\rangle$, $|2\rangle$, $|3\rangle$ and 
$|4\rangle$ respectively. It is convenient to introduce 
the reference state 
\begin{equation}
|vac\rangle=|4\rangle_1\otimes_1\cdots\otimes_s
|4\rangle_L
\end{equation}
Using the explicit expression of the L-operator, one can find
that the monodromy matrix acting on the reference state
takes the form
\begin{equation}
\ct(\mu)|vac\rangle=\left(\begin{array}{cccc}
A_1(\mu) & 0 & 0 & 0\\[3mm]
\ct_{21}(\mu)& A_2(\mu) & 0 & 0\\[3mm]
\ct_{31}(\mu)& 0& A_3(\mu) & 0 \\[3mm]
\ct_{41}(\mu) & \ct_{42}(\mu) & \ct_{43}(\mu) 
& A_4(\mu) \end{array} \right) |vac\rangle
 \end{equation}
where
\begin{eqnarray*}
A_4(\mu)&=&[-\alpha^2(\mu)e^{h(\mu)}]^L\\[5mm]
A_2(\mu)&=&A_3(\mu)=[\alpha(\mu)\gamma(\mu)e^{-h(\mu)}]^L\\[5mm]
A_1(\mu)&=&[-\gamma^2(\mu)e^{h(\mu)}]^L
\end{eqnarray*}

  Unlike the case where the nested Bethe Ansatz method is
applicable, the operator $\ct_{21}$ is related to operator
$\ct_{43}$, creating an electron with spin-down. Generally,
the  relation is very complicated. When they act on the reference
state, however, the situation becomes  simple. An algebraic 
calculation shows         
\begin{equation}
\ct_{21}(\tilde{\mu})|vac\rangle=\frac{(-1)^L\gamma^{2L-1}
   (\tilde{\mu})e^{(L-1)h(\tilde{\mu})}}
  {\gamma^{L-1}\alpha(\mu)^Le^{-(L-1)h(\mu)}}
 \ct_{43}(\mu)|rac\rangle
\end{equation}
where the quantity with tilde is defined by
\begin{equation}
e^{-2h(\tilde{\mu})}\frac{\alpha(\tilde{\mu})}
{\gamma(\tilde{\mu})}=
e^{2h(\mu)}\frac{\alpha(\mu)}
{\gamma(\mu)}
\end{equation}
Further, one can show 
$$\ct_{21}(\tilde{\mu})\ct_{43}(\mu_1)\cdots \ct_{43}(\mu_n)|vac
\rangle\propto  \ct_{43}(\mu)\ct_{43}(\mu_1)\cdots 
\ct_{43}(\mu_n)|vac\rangle. 
$$
It is worthy to notice that this relation is valid only for
the reference state; when acting on the other states, 
it will be
invalid. There is a similar relation between $\ct_{31}$ and 
$\ct_{42}$. Due to this relation, we can construct the special
states with all spin-up (spin-down) with $\ct_{43}$ ($\ct_{42}$) 
$$|\Psi_N\rangle=\ct_{43}(\mu_1)\cdots \ct_{43}(\mu_N)
|\Psi_N\rangle$$
Using the commutative relations, we can find
\begin{eqnarray*}
t_H(\mu)|\Psi_N\rangle&=&\Lambda(\mu)|\Psi_N\rangle
     + \mbox{unwanted terms}.
\end{eqnarray*}
where
\begin{eqnarray}
\Lambda(\mu)
&=&\displaystyle
     A_4(\mu)\prod_{j=1}^N\frac{\rho_1(\mu_j,\mu)}
     {i\rho_9(\mu_j,\mu)} -
     A_3(\mu)\prod_{j=1}^N\frac{-\rho_4(\mu,\mu_j)}
     {i\rho_9(\mu,\mu_j)}\nonumber \\
& &\displaystyle
     -A_2(\mu)\prod_{j=1}^N\frac{i\rho_{10}(\mu,\mu_j)}
     {\rho_1(\mu,\mu_j)-\rho_3(\mu,\mu_j)}\nonumber\\
& &\displaystyle
     +A_1(\mu)\prod_{j=1}^N\frac{-i\rho_{10}(\mu,\mu_j)}
     {\rho_1(\mu,\mu_j)-\rho_3(\mu,\mu_j)}
\end{eqnarray}
The  vanishing unwanted terms gives the Bethe Ansatz equation
\begin{equation}
[-e^{2h(\mu_j)}\frac{\alpha(\mu_j)}{\gamma(\mu_j)}]^L=1
\end{equation}
Thus, the states $|\Psi_N\rangle$ are really the eigenstates of 
the transfer matrix $t(\mu)$ if the spectrum parameters are 
appropriately chosen to satisfy the Bethe Ansatz equation (20). 

The general states with $N$ spin-down and $N-M$ spin-up are 
given by sums of products of the combination of $\ct_{21}$, 
$\ct_{31}$, and $\ct_{4j}, j=1,2,3$ acting 
on the reference state. Let us consider a special case 
$N-M=M=1$. The general form is
\begin{eqnarray}
|\Psi_{1,1}\rangle
&=&\left\{f_1(\mu_1,\mu_2)\ct_{42}(\mu_1)\ct_{43}(\mu_2)
          +f_2(\mu_1,\mu_2)\ct_{43}(\mu_1)\ct_{42}(\mu_2)
       \right.        \nonumber\\[3mm]
& &+\;f_3(\mu_1,\mu_2)\ct_{21}(\mu_1)\ct_{31}(\mu_2)+
      f_4(\mu_1,\mu_2)\ct_{31}(\mu_1)\ct_{21}(\mu_2)
                                    \nonumber\\[3mm]
& &\left.+\;f_5(\mu_1,\mu_2)\ct_{41}(\mu_1)+
   f_6(\mu_1,\mu_2)\ct_{41}(\mu_2)\right\}|vac\rangle
\end{eqnarray}
We can determine the all coefficients $f_j$ by requiring 
the r.h.s. of equation (21) to be the eigenvector of the 
transfer matrix.  The general solution is very complicated. 
Fortunately, we  can show by an explicit calculation that 
$|\Psi_{1,1}\rangle$ is the eigenstate if $f_3=f_4=0$ with  
appropriate $f_5$ and $f_6$. Using the Super-Yang-Baxter 
relation, we find the eigenvalue for the case when $f_1=-f_2$
\begin{eqnarray} \Lambda(\mu) 
&=&\displaystyle
     A_4(\mu)\prod_{j=1}^2\frac{\rho_1(\mu_j,\mu)}
     {i\rho_9(\mu_j,\mu)} 
     +A_1(\mu)\prod_{j=1}^N\frac{-i\rho_{10}(\mu,\mu_j)}
     {\rho_1(\mu,\mu_j)-\rho_3(\mu,\mu_j)}\nonumber \\[3mm]
& &\displaystyle-A_2(\mu)
   \left\{\frac{\rho_4(\mu,\mu_1)\rho_{10}(\mu,\mu_2)}
        {\rho_9(\mu,\mu_1)[\rho_1(\mu,\mu_2)-\rho_3(\mu,\mu_2)]}
    \right. \nonumber\\[3mm]
& &\displaystyle\;+ \frac{\rho_4(\mu,\mu_2)\rho_{10}(\mu,\mu_1)}
           {\rho_9(\mu,\mu_2)[\rho_1(\mu,\mu_1)-\rho_3(\mu,\mu_1)]} 
    \nonumber\\[3mm] 
& &\displaystyle\;-
\left[\frac{\rho_{10}(\mu,\mu_1)}{\rho_1(\mu,\mu_1)-\rho_3(\mu,\mu_1)}
     +\frac{\rho_4(\mu,\mu_1)}{\rho_9(\mu,\mu_1)}\right]
      \nonumber\\[3mm]
& &\displaystyle\;\;\times \left.
\left[\frac{\rho_{10}(\mu,\mu_2)}{\rho_1(\mu,\mu_2)-\rho_3(\mu,\mu_2)}
      +\frac{\rho_4(\mu,\mu_2)}{\rho_9(\mu,\mu_2)}
    \right]\right\} \label{21spe}
\end{eqnarray}
The  vanishing unwanted terms gives the Bethe Ansatz equation
\begin{equation}
[\prod_{j=1}^2\frac{-\alpha(\mu_j)}{\gamma(\mu_j)}
e^{2h(\mu_j)}]^L=1
\end{equation}
For the case  $f_1=f_2$ the eigenvalue and the Bethe Ansatz 
equations are given by  the equations (19) and (20), 
respectively.  It is worthy to point out that these states 
considered here  are not complete. For example, 
consider the case $f_1=f_2=0$, one can get similar results.

\section{Analytic Bethe Ansatz for the Hubbard model}

  In the last  section, we applied the algebraic Bethe Ansatz
method to some eigenstates of the Hubbard model. For
general states, the straightforward calculation becomes very 
complicated. In this section, however, we want to discuss an  
analytic Bethe Ansatz method to the same problem based on  
the hints given by the above results. We should generalize 
the standard ABA \cite{Res1,VR,Res2,Mar} in which the crossing 
symmetry and asymptotic behaviour play a key role. We can not
apply the same argument to the eigenvalue of the Hubbard 
model in which there is no such crossing symmetry for the 
R matrix.

Let us first investigate  the special solutions and show how 
to generalize it from the viewpoint of ABA. One may 
understand that the analytic property of function 
$\Lambda(\mu)$ leads to the Bethe Ansatz equation (20).
It is clear that $\rho_9(\mu,\mu_j)=0$ is the simple pole of 
$\Lambda(\mu)$ (equation (19)). In order to keep the analytic 
property  of the eigenvalue, the residue at such
pole must be zero. The Bethe Ansatz equation is nothing 
but the condition of vanishing residue. Similarly the 
vanishing residue at the pole 
$\rho_1(\mu,\mu_j)=\rho_3(\mu,\mu_j)$ gives the same Bethe 
Ansatz equation. This property can be generalized to all 
kinds of states with different particles. The eigenvalue 
function should be analytic and has only superficial 
simple poles. The vanishing residues at such poles will give
the Bethe Ansatz equations. 

Let us discuss the general eigenvalue of  the Hubbard model. 
The special solutions (19) and (22) together with some standard
knowledge of algebraic Bethe Ansatz contain enough 
message about the general one. It consists of four terms 
which are proportional to $A_j(\mu),j=1,2,3,4$ respectively.  
The terms involving $A_1$ and $A_4$ are  dependent only on 
the total number of electrons. The other terms depend on 
both the total number $N$ of electrons and the number $M$ 
of spin-up electrons as shown in the last section. Thus, 
the general eigenvalue should be 
\begin{eqnarray}
\Lambda(\mu)&=&\displaystyle
              A_4(\mu)\prod_{j=1}^N\frac{\rho_1(\mu_j,\mu)}
                 {i\rho_9(\mu_j,\mu)}
              -A_3(\mu)\prod_{j=1}^N\frac{-\rho_4(\mu,\mu_j)}
                 {i\rho_9(\mu_j,\mu)}\prod_{m=1}^Mg_3(\mu,\lambda_m)
                \nonumber\\
             & &\displaystyle 
                -A_2(\mu)\prod_{j=1}^N\frac{-i\rho_{10}(\mu,\mu_j)}
               {\rho_1(\mu,\mu_j)-\rho_3(\mu,\mu_j)}
                \prod_{m=1}^Mg_2(\mu,\lambda_m)\nonumber\\
             & &\displaystyle
              +A_1(\mu)\prod_{j=1}^N\frac{-i\rho_{10}(\mu,\mu_j)}
               {\rho_1(\mu,\mu_j)-\rho_3(\mu,\mu_j)}
\end{eqnarray}
where $g_2$ and $g_3$ are undetermined functions. The $\mu_j$ and 
$\lambda_m$ are free parameters. $N$ is the total number of electrons, 
$M$ the number of spin-up electrons. 

Now, we show how  the analytic property of the 
eigenvalue restricts the unknown functions. 
First, $\Lambda(\mu)$ has two sets of poles 
related to parameters $\mu_j$. One (case A)
is controlled by the null denominator of the first 
two terms of equation (24).
 The another (case B) is from the last two terms.
For case A the position of poles is determined by
\begin{equation}
e^{2h(\mu)}\frac{\alpha(\mu)}{\gamma(\mu)}
=e^{2h(\mu_j)}\frac{\alpha(\mu_j)}{\gamma(\mu_j)}
\end{equation}
Due to the $i\pi$-period of $h(\mu)$, we can get $\mu=\mu_j$
in the region $0\le \mu_j\le \pi$. At these poles, the functions
$\rho_1(\mu,\mu_j)-\rho_3(\mu,\mu_j)$ and 
$\rho_{10}(\mu,\mu_j)$ also vanish, But, the ratio is finite. 
So, the singularity at these poles is dominated
by the terms in the first line. The null residue requires 
\begin{equation}
\left[-e^{2h(\mu_j)}\frac{\alpha(\mu_j)}{\alpha(\mu_j)}\right]^L
=\prod_{m=1}^Mg_3(\mu_j,\lambda_m) \label{pol1}
\end{equation}

For case B, the position of poles satisfies
\begin{equation}
\begin{array}{rcl}
0&=&[e^{-h(\mu_j)-h(\mu)}\gamma(\mu_j)\alpha(\mu_j)-
e^{h(\mu_j)+h(\mu)}\gamma(\mu)\alpha(\mu)]\\
& &\displaystyle \times
[e^{h(\mu)-h(\mu_j)}\gamma(\mu)\alpha(\mu_j)-
e^{h(\mu_j)-h(\mu)}\alpha(\mu)\gamma(\mu_j)]
\end{array}
\end{equation}
The first part is equal to $\rho_9(\mu,\mu_j)=0$. It is not a
pole due to $\rho_{10}(\mu,\mu_j)=0$ at this point. 
The real pole locates in $\tilde{\mu}_j$  
\begin{equation}
e^{-2h(\tilde{\mu_j})}\frac{\alpha(\tilde{\mu_j})}
{\gamma(\tilde{\mu_j})}
=e^{2h(\mu_j)}\frac{\alpha(\mu_j)}{\gamma(\mu_j)}\label{pol2}
\end{equation}
The vanishing residue at $\mu=\tilde{\mu}_j$ gives
\begin{equation}
\left[-e^{2h(\tilde{\mu_j})}\frac{\gamma(\tilde{\mu_j})}
{\alpha(\tilde{\mu_j})}\right]^L=
\prod_{j=1}^Ng_2(\mu_j,\lambda_m)
\end{equation}
In order to keep the analytic property of the eigenvalue, 
Equations (\ref{pol1}) and (\ref{pol2}) must be satisfied 
simultaneously, which leads to the following functional
relation 
\begin{equation} \prod_{m=1}^Mg_3(\mu_j,\lambda_m)
=\prod_{m=1}^Mg^{-1}_2(\tilde{\mu}_j,\lambda_m)
\end{equation}
We find that it is convenient to introduce the new variables 
$k$ and $k_j$
\begin{equation}
e^{ik}=-e^{2h(\mu)}\frac{\alpha(\mu)}{\gamma(\mu)}
\end{equation}
In terms of the new variables the relation between the 
quantities with and without tilde is very simple
\begin{equation}
{\sin}(\tilde{k}_j) = \sin(k_j)+i\frac{U}2
\end{equation}

At this stage,  we need to know some properties of the 
undetermined functions $g_i$ which hides already in the 
special solution with $N=2,M=1$.  In terms of $k_j$, 
equation (\ref{21spe}) changes into
\begin{eqnarray}
\Lambda(\mu)&=&\left(-\alpha^2(\mu)e^{h(\mu)}\right)^L
               \Lambda(k)\nonumber \\
\Lambda(k)&=&\displaystyle\prod_{j=1}^2
             \frac{2i\cos((k+k_j)/2)\cos((\tilde{k}+k_j)/2)}
              {i\sin (k)-i\sin (k_j)} \nonumber \\
          & &\displaystyle-\; e^{-ikL}\prod_{j=1}^N
             \frac{2i\cos((k+k_j)/2)\cos((\tilde{k}+k_j)/2)}
              {i\sin (k)-i\sin (k_j)}\nonumber\\
          & & \displaystyle \times
              \frac{i2\sin (k)-i\sin(K_1)-i\sin(k_2)-U/2}
              {i2\sin (k)-i\sin(k_1)-i\sin(k_2)+U/2} \nonumber \\
          & &\displaystyle-\; e^{-ikL}\prod_{j=1}^2
             \frac{2i\cos((k+k_j)/2)\cos((\tilde{k}+k_j)/2)}
              {i\sin (k)-i\sin (k_j)+U/2}\nonumber \\
	  & &\displaystyle \times
             \frac{i2\sin (k)-i\sin(K_1)-i\sin(k_2)+3U/2}
              {i2\sin (k)-i\sin(k_1)-i\sin(k_2)+U/2} \nonumber \\
          & &\displaystyle-\; e^{-i(k+\tilde{k})L}\prod_{j=1}^2
             \frac{2i\cos((k+k_j)/2)\cos((\tilde{k}+k_j)/2)}
              {i\sin (k)-i\sin (k_j)+U/2} 
\end{eqnarray}
It is clear  that the  eigenvalue has another simple pole
$2\sin(k)=\sin(k_1)+\sin(k_2)-U/2$. The Bethe Ansatz 
equation ensures the analytic property of the eigenvalue. 
This strongly suggests that the undetermined functions have 
the simple poles of the form  $\sin (k)=const$. On the other 
hand, the analytic property of the eigenvalue requires that 
$g_2$ and $g_3$  must have same poles. Therefore, the general 
form of $g_2(\mu,\lambda)$ and $g_3(\mu,\lambda)$ is
\begin{equation}
g_2(\mu,\lambda)=\frac{P_2(k,\lambda)}{i\sin (k)-\lambda+U/4}
,g_3(\mu,\lambda)=\frac{P_3(k,\lambda)}{i\sin (k)-\lambda+U/4}
\end{equation}
where the function $P_2(k,\lambda)$ and $P_3(k,\lambda)$
are integral  functions. The general form is $P_r(k,\lambda)=
\sum_{n=0}a^r_n(k)(\lambda)^n$. Substituting it into equation 
(30), we have found
\begin{equation}
\begin{array}{rcl}
a^2_0(k)&=&[i\sin (k)-U/4]a^2_1(k)\\[3mm]
a^3_0(k)&=&[i\sin (k)+3U/4](a^3_1(k)\\[3mm]
a^2_n(k)&=&a^2_n(k)=0, n\geq 2 \\[3mm]
a^3_1(k)&=[a^2_1(\hat{k})]^{-1}
\end{array}
\end{equation}
where $\hat{k}$ is defined by $\sin( \hat{k})=\sin (k)-iU/2$.
Moreover, the function $a^2_1(k)$ is analytic and has no 
zero in the complex plain. It will be fixed by the asymptotic 
behaviour of the transfer matrix. Let us assume $U\le 0$ and 
$\mu\longrightarrow -i\infty$, then the eigenvalue approaches to
\begin{equation}
\Lambda(\mu)\longrightarrow e^{3L\infty}\{[a^2_1(\infty)]^M
+[a^2_1(\infty)]^{-M}\}.
\end{equation}
Comparing this with the asymptotic behaviour of $t(\mu)
\longrightarrow e^{3L\infty}$, we get $a^2_1(\infty) $ 
is a no-zero constant. Based upon the knowledge of analysis 
such as Liouville's theorem on integral functions, we 
arrive at $a^2_1(k)$ being a no-zero constant. A special form 
of $\Lambda(\mu)$ under $N=M=1$ fixes this constant to be unit. 
Finally, we arrive at the final results
\begin{eqnarray}
\Lambda(\mu)&=&\left(-\alpha^2(\mu)e^{h(\mu)}\right)^L
               \Lambda(k)\nonumber \\
\Lambda(k)&=&\displaystyle\prod_{j=1}^N
             \frac{2i\cos((k+k_j)/2)\cos((\tilde{k}+k_j)/2)}
              {i\sin (k)-i\sin (k_j)} \nonumber \\
          &&\displaystyle-\; e^{-ikL}\prod_{j=1}^N
             \frac{2i\cos((k+k_j)/2)\cos((\tilde{k}+k_j)/2)}
              {i\sin (k)-i\sin (k_j)}\nonumber\\
   & & \displaystyle \times
             \prod_{m=1}^M\frac{i\sin (k)-\lambda_m-U/4}
              {i\sin (k)-\lambda_m+U/4} \nonumber \\
          \nonumber \\
         &&\displaystyle-\; e^{-ikL}\prod_{j=1}^N
             \frac{2i\cos((k+k_j)/2)\cos((\tilde{k}+k_j)/2)}
              {i\sin (k)-i\sin (k_j)+U/2}\nonumber \\
	      & &\displaystyle \times
             \prod_{m=1}^M\frac{i\sin (k)-\lambda_m+3U/4}
              {i\sin (k)-\lambda_m+U/4} \nonumber \\
           &&\displaystyle-\; e^{-i(k+\tilde{k})L}\prod_{j=1}^N
             \frac{2i\cos((k+k_j)/2)\cos((\tilde{k}+k_j)/2)}
              {i\sin (k)-i\sin (k_j)+U/2} \label{hueigen} 
\end{eqnarray}
The parameters satisfy the following Bethe Ansatz equations
\begin{eqnarray}
e^{ik_jL}&=&\displaystyle\prod_{m=1}^M
             \frac{i\sin(k_j)-\lambda_m-U/4}
                  {i\sin(k_j)-\lambda_m+U/4}\nonumber \\
-\displaystyle\prod_{m=1}^M\frac{\lambda_r-\lambda_m-U/2}
                  {\lambda_r-\lambda_m+U/2}&=&
\displaystyle\prod_{j=1}^N
             \frac{i\sin(k_j)-\lambda_r+U/4}
                  {i\sin(k_j)-\lambda_m-U/4}\label{huba}
\end{eqnarray}
Differentiating $\log (\Lambda(\mu))$ at $ \mu=\pi/4$ will give
the energy of the Hubbard model, which coincides with the 
one given in \cite{LW}
\begin{equation}
E=\frac{UL}{4}-\frac{NU}2-\sum_{j=1}^N\cos (k_j)
\end{equation}
We have checked that equations (\ref{hueigen}) and (\ref{huba})
with $N=2,M=1$ coincide with the result obtained in section 3.

\section{ABA for the coupled XY model}

In this section we investigate the eigenvalue of the transfer
matrix of the coupled XY model with twisted boundary condition. 

The L-operator related to the coupled XY model is \cite{Sha1,Sha2}:
\begin{eqnarray}
\lefteqn{L_m(\mu)=}\nonumber \\ & &
\left(\begin{array}{cccc}
e^{h(\mu)}p_m^+q^+_m &p_m^+\tau^-_m &\sigma^-_mq^+_m
                    &e^{h(\mu)}\sigma^-_m\tau^-_m\\[3mm]
p_m^+\tau^+_m &e^{-h(\mu)}p_m^+q^-_m
    &e^{-h(\mu)}\sigma^-_m\tau^+_m &\sigma^-_mq^-_m\\[3mm]
\sigma^+_mq^+_m &e^{-h(\mu)}\sigma^+_m\tau^-_m
    &e^{-h(\mu)}p_m^-q^+_m &p_m^-\tau^-_m \\[3mm]
e^{h(\mu)}\sigma^+_m\tau^+_m  &\sigma^+_mq^-_m
    &p_m^-\tau^+_m &e^{h(\mu)}p_m^-q^-_m
\end{array}\right) 
\end{eqnarray}
where $\sigma^a_m$ and $\tau^a_m$ are two independent Pauli matrices 
located in $m$-th site. The operator $ p^{\pm}$ and $q^{\pm}$
read
\begin{equation}
\begin{array}{rcl}
p^{\pm}_m&=&w_4(\mu)\pm w_3(\mu)\sigma^z_m,\\[3mm]
q^{\pm}_m&=&w_4(\mu)\pm w_3(\mu)\tau^z_m.
\end{array}
\end{equation}
In references \cite{Sha1,Sha2}, it was shown that this L-operator 
satisfies the Yang-Baxter equation
\begin{equation}
R(\mu,\nu)L_m(\mu)\otimes L_m(\nu)=
L_m(\nu)\otimes L_m(\mu) R(\mu,\nu)
\end{equation}
The R matrix is
\begin{equation}\left(
\begin{array}{cccccccccccccccc}
\rho_1&0&0&0    &0&0&0&0         &0&0&0&0          &0&0&0&0 \\
0&\rho_2&0&0    &\rho_9&0&0&0   &0&0&0&0          &0&0&0&0 \\
0&0&\rho_2&0    &0&0&0&0         &\rho_9&0&0&0    &0&0&0&0 \\
0&0&0&\rho_3    &0&0&\rho_6&0  &0&\rho_6&0&0  
                              &-\rho_8&0&0&0 \\
0&\rho_{10}&0&0  &\rho_2&0&0&0  &0&0&0&0         &0&0&0&0 \\
0&0&0&0       &0&\rho_6&0&0     &0&0&0&0          &0&0&0&0 \\
0&0&0&\rho_6  &0&0&\rho_5&0   &0&-\rho_7&0&0  
                               &\rho_6&0&0&0 \\
0&0&0&0    &0&0&0&\rho_2   &0&0&0&0    &0&\rho_{10}&0&0 \\
0&0&\rho_{10}&0  &0&0&0&0    &\rho_2&0&0&0     &0&0&0&0  \\
0&0&0&\rho_6    &0&0&-\rho_7&0  &0&\rho_5&0&0  
                                        &\rho_6&0&0&0 \\
0&0&0&0         &0&0&0&0      &0&0&\rho_4&0     &0&0&0&0 \\
0&0&0&0         &0&0&0&0      &0&0&0&\rho_2     
                               &0&0&\rho_{10}&0 \\
0&0&0&-\rho_8    &0&0&\rho_6&0  &0&\rho_6&0&0  
                                       &\rho_3&0&0&0 \\
0&0&0&0    &0&0&0&\rho_9   &0&0&0&0    &0&\rho_2&0&0 \\
0&0&0&0         &0&0&0&0      &0&0&0&\rho_9     
                               &0&0&\rho_2&0 \\
0&0&0&0         &0&0&0&0      &0&0&0&0  &0&0&0&\rho_{1} 
\end{array} \right)
\end{equation}
This Yang-Baxter relation ensures the monodromy matrix 
$T(\mu)=L_L(\mu)\otimes \cdots\otimes L_1(\mu)$ satisfying the 
Yang-Baxter relation. 
In order to simplify our calculation, we choose the ferromagnetic
state ( all spin-down states) as the reference state. From the explicit
expression of the L-operator, we have
\begin{equation}
T(\mu)|vac\rangle=\left(\begin{array}{cccc}
A_1(\mu) & 0 & 0 & 0\\[3mm]
T_{21}(\mu)& A_2(\mu) & 0 & 0\\[3mm]
T_{31}(\mu)& 0& A_3(\mu) & 0 \\[3mm]
T_{41}(\mu) & T_{42}(\mu) & T_{43}(\mu) & A_4(\mu) \end{array} \right)
 |vac\rangle
 \end{equation}
where
\begin{eqnarray}
A_4(\mu)&=&[\alpha^2(\mu)e^{h(\mu)}]^L\nonumber \\[5mm]
 A_2(\mu)&=&A_3(\mu)=[\alpha(\mu)\gamma(\mu)e^{-h(\mu)}]^L
       \nonumber\\[5mm]
 A_1(\mu)&=&[\gamma^2(\mu)e^{h(\mu)}]^L
\end{eqnarray}
Similarly, one can define the transfer matrix $t(\mu)=st T(\mu)$ and 
find the eigenvalue of it, which will related to the periodic 
boundary condition. The eigenvalue of the diagonal-of-diagonal 
transfer matrix of  this model with periodic condition was 
found by Bariev \cite{Bar} in terms of the coordinate Bethe 
Ansatz method. 

In order to consider twisted boundary 
condition, we introduce the generalized transfer matrix
\begin{equation}
t^g(\mu)=T_{11}(\mu)a^{i\beta_1}+T_{22}(\mu)a^{i\beta_2}+
         T_{33}(\mu)a^{i\beta_3}+T_{44}(\mu)a^{i\beta_4}
\end{equation}
where 
\begin{equation}
\begin{array}{rcl}
a^{i\beta_1}&=& e^{a_{\sigma} N_{\sigma}+
                a_{\tau} N_{\tau}+a_0} \\[3mm]
a^{i\beta_2}&=& e^{c_{\sigma} N_{\sigma}+
                c_{\tau} N_{\tau}+c_0} \\[3mm]
a^{i\beta_3}&=& e^{-i\beta_2}\\[3mm]
a^{i\beta_4}&=& e^{-i\beta_1}
\end{array}
\end{equation}
where $a_s$ and $c_s$ are free parameters, $N_{\sigma}$ 
($N_{\tau}$) the total number of $\sigma$-spin 
($\tau$-spin). Now, we want to find  the eigenvalue  
of $t^g(\mu)$ by means of the analytic Bethe Ansatz method. 
First, by using  the algebraic Bethe Ansatz method, we find 
the eigenvalue of the states with $N$ $\tau$-spin 
(or $\sigma$-spin) flipping from the reference state. 
After a  long but direct calculation, we arrive at
\begin{equation}
\begin{array}{rcl}
\Lambda_N(\mu)&=&\displaystyle \epsilon\; e^{a_s N+a_0}
                 [e^{h(\mu)}\alpha^2(\mu)]^L \prod_{j=1}^N
                 \frac{\rho_1(\mu_j,\mu)}{\rho_9(\mu_j,\mu)}\\[3mm]
& &+ \displaystyle \epsilon'\; e^{c_s N+c_0}
                 [e^{-h(\mu)}\alpha(\mu)\gamma(\mu)]^L \prod_{j=1}^N
                 \frac{\rho_4(\mu,\mu_j)}{\rho_9(\mu,\mu_j)}\\[3mm]
& &+\displaystyle \epsilon'\; e^{-c_sN-c_0}
                 [e^{-h(\mu)}\alpha(\mu)\gamma(\mu)]^L \prod_{j=1}^N
                 \frac{\rho_{10}(\mu,\mu_j)}{\rho_1(\mu_j,\mu)-
                 \rho_3(\mu,\mu_j)}\\[3mm]
& &+\displaystyle \epsilon\; e^{-a_sN-a_0}
                 [e^{h(\mu)}\gamma^2(\mu)]^L \prod_{j=1}^N
                 \frac{\rho_1(\mu,\mu_j)}{\rho_3(\mu,\mu_j)-\rho_1(
                 \mu,\mu_j)}
\end{array}
\end{equation}
where $s=\sigma$ ($s=\tau$) for all $\sigma$ ($\tau$)  spin-up states,
the parameters $\mu_j$ are determined by 
\begin{equation}
\left[e^{2h(\mu_j)}\frac{\alpha(\mu_j)}
   {\gamma(\mu_j)}\right]=(-1)^{N+1}
e^{(c_s-a_s)N+c_0-a_0}
\end{equation}
>From this expression and the similar arguments in the last
section, we can write the 
general form of the eigenvalue
\begin{equation}
\begin{array}{rcl}
\Lambda(\mu)&=&\displaystyle  e^{a_0+a_{\sigma}M
                +a_{\tau}(N-M)}
                 [e^{h(\mu)}\alpha^2(\mu)]^L \prod_{j=1}^N
                 \frac{\rho_1(\mu_j,\mu)}{\rho_9(\mu_j,\mu)}\\[3mm]
& &+ \displaystyle  e^{c_0+c_{\sigma}M
                +c_{\tau}(N-M)}
                 [e^{-h(\mu)}\alpha(\mu)\gamma(\mu)]^L \\[3mm]
& &\displaystyle\times \prod_{j=1}^N
                 \frac{\rho_4(\mu,\mu_j)}{\rho_9(\mu,\mu_j)}
                 \prod_{m=1}^Mg_3(\mu,\lambda_m)\\[3mm]
& &+\displaystyle  e^{-c_0-c_{\sigma}M-c_{\tau}(N-M)}
                 [e^{-h(\mu)}\alpha(\mu)\gamma(\mu)]^L \\[3mm]
& &\displaystyle \times \prod_{j=1}^N
                 \frac{\rho_{10}(\mu,\mu_j)}{\rho_1(\mu_j,\mu)-
                 \rho_3(\mu,\mu_j)}
                 \prod_{m=1}^Mg_2(\mu,\lambda_m)\\[3mm]
& &+\displaystyle \epsilon \epsilon e^{-a_0-a_{\sigma}M
                -a_{\tau}(N-M)}
                 [e^{h(\mu)}\gamma^2(\mu)]^L \prod_{j=1}^N
                 \frac{\rho_1(\mu,\mu_j)}{\rho_3(\mu,\mu_j)-\rho_1(
                 \mu,\mu_j)}
\end{array}
\end{equation}
Second, consider the singularity of the $\Lambda(\mu)$ at the poles 
related to the parameters $\mu_j$.  As done in the Hubbard model, 
the null residue
condition requires the following relation
\begin{equation}
\prod_{m=1}^Mg_3(\mu,\lambda_m)=\prod_{m=1}^Mg^{-1}_2(\tilde{\mu}
       ,\lambda_m)
\end{equation}
This equation is same as that in the Hubbard model (equation (34)). So, we 
can use the results of the Hubbard model. 
\begin{equation}
\begin{array}{rcl}
g_2(\mu,\lambda)&=&\displaystyle
        c\frac{i\sin (k)-\lambda-U/4}{i\sin (k)-\lambda+U/4}\\[3mm]
g_3(\mu,\lambda)&=&\displaystyle
        \frac1c\frac{i\sin (k)-\lambda-U/4}{i\sin (k)-\lambda+U/4}
\end{array}
\end{equation}
here we have used  the same definition of $k$ as one in the Hubbard 
model. One should note that in this  case, the constant $c$ in the 
above equation not being  $1$ as in the Hubbard model. 
Taking $N=M=1$ in equation(50) and comparing with 
equation (48), one can get $c=-1$. Thus, we obtain the final results
\begin{equation}
\begin{array}{rcl}
\Lambda(\mu)&=&[e^{h(\mu)}\alpha^2(\mu)]^L\Lambda(k)\\[5mm]
\Lambda(k)&=&\displaystyle (-1)^N
    e^{a_0+a_{\sigma}M +a_{\tau}(N-M)} \prod_{j=1}^N
                 \frac{2\cos((k+k_j)/2)\cos((\tilde{k}+k_j)/2)}
		  {i\sin(k)-i\sin(k_j)}\\[5mm]
& &+ \displaystyle  e^{c_0+c_{\sigma}M+c_{\tau}(N-M)}
     \prod_{j=1}^N\frac{2\cos((k+k_j)/2)\cos((\tilde{k}+k_j)/2)}
		  {i\sin(k)-i\sin(k_j)}\\[5mm]
& & \displaystyle \;\times(-1)^{L+M}e^{-ikL}
\prod_{m=1}^M\frac{i\sin (k)-\lambda_m-U/4}
                 {i\sin (k)-\lambda_m+U/4}\\[5mm]
& &+\displaystyle  e^{-c_0-c_{\sigma}M-c_{\tau}(N-M)}
    \prod_{j=1}^N\frac{2\cos((k+k_j)/2)\cos((\tilde{k}+k_j)/2)}
            {i\sin (k)-i\sin(k_j)+U/2}\\[5mm]
& &\displaystyle \;\times (-1)^{L+M+N}e^{-ikL}
	  \prod_{m=1}^M\frac{i\sin(k)-\lambda_m+3U/4}
                {i\sin(k)-\lambda_m+U/4}\\[3mm]
& &+\displaystyle  e^{-a_0-a_{\sigma}M
                -a_{\tau}(N-M)}e^{-i(k+\tilde{k})L}
    \prod_{j=1}^N\frac{2\cos((k+k_j)/2)\cos((\tilde{k}+k_j)/2)}
            {i\sin (k)-i\sin(k_j)+U/2}
\end{array}
\end{equation}
The Bethe Ansatz equations are 
\begin{equation}
\begin{array}{rcl}
(-1)^{M+N+1+L}e^{ik_jL}
&= &\displaystyle e^{c_0-a_0+(c_{\sigma}-a_{\sigma})M+
    (c_{\tau}-a_{\tau})(N-M)} \\[3mm]
& &\displaystyle\; \times \prod_{m=1}^M
   \frac{i\sin(k_j)-\lambda_m-U/4}
   {i\sin(k_j)-\lambda_m+U/4}\\[5mm]
\displaystyle\prod_{j=1}^N \frac{i\sin(k_j)-\lambda_r+U/4}
                  {i\sin(k_j)-\lambda_r-U/4}
&=&\displaystyle (-1)^{N+1}e^{2(c_0+c_{\sigma}M+
                 c_{\tau}(N-M))}\\[3mm]
& &\displaystyle\;\times \prod_{m=1}^M\frac{\lambda_r-
 \lambda_m-U/2}{\lambda_r-\lambda_m+U/2}
\end{array}
\end{equation}
These are the exact solution of the coupled XY model with 
twisted boundary condition. It is of  interest that it 
recovers the results of the Hubbard model when  
$a_{\sigma}=a_{\tau}=c_{\sigma}=-c_{\tau}=-i\pi/2$ and 
$a_0=iL\pi, c_0=i\pi$. This gives a precise relation 
between the two models and make it clear why the extra 
factor appears among the Bethe Ansatz equations by Lieb
and Wu \cite{LW} and by Shastry \cite{Sha3}. 
When $a_s=c_s=0, s=\sigma,\tau,0$, 
they reduced into the periodic case. The correspondence 
between our notation and that in \cite{Sha3} is 
$2i\sin(k_j)=z_j^{-1}-z_j$. 

\section{Twisted boundary condition}

In this section, we will discuss the boundary condition related 
to our generalized transfer matrix (46) and prove the equality
of the periodic Hubbard model to  the twisted coupled XY model
in terms of the transfer matrices and the hamiltonians.

First, we derive the hamiltonian related to the $t^g$ (46)
by using the standard  method. After a straightforward 
calculation, we arrive at 
\begin{equation}
\begin{array}{rcl}
H&=&\displaystyle\sum_{m=1}^{L-1}(\sigma^+_{m+1}a^-_{m}
    +\sigma^+_{m}\sigma^-_{m+1}+\tau^+_{m+1}\tau^-_{m}
    +\tau^+_{m}\tau^-_{m+1})+
    \frac{U}4\sum_{m=1}^{L}\sigma^z_m\tau^z_m\\[3mm]
&&\displaystyle +\exp\{-\epsilon[a_0+c_0+(a_{\sigma}
      +c_{\sigma})N_{\sigma}+(a_{\tau}+c_{\tau})N_{\tau}]\}
      \sigma^+_N\sigma^-_1 \\[3mm]
&&\displaystyle +\exp\{\epsilon[a_0+c_0+(a_{\sigma}
      +c_{\sigma})N_{\sigma}+(a_{\tau}+c_{\tau})N_{\tau}]\}
      \sigma^-_N\sigma^+_1 \\[3mm]
&&\displaystyle +\exp\{-\epsilon'[a_0-c_0+(a_{\sigma}
      -c_{\sigma})N_{\sigma}+(a_{\tau}-c_{\tau})N_{\tau}]\}
      \tau^+_N\tau^-_1 \\[3mm]
&&\displaystyle +\exp\{\epsilon'[a_0-c_0+(a_{\sigma}
      -c_{\sigma})N_{\sigma}+(a_{\tau}-c_{\tau})N_{\tau}]\}
      \tau^-_N\tau^+_1 
\end{array}
\end{equation}
where 
\begin{equation}
\begin{array}{cc}
\epsilon=\left\{\begin{array}{cc}
 1& \tau \mbox{spin up}\\ -1& \tau \mbox{spin down}
  \end{array} \right. ,&
\epsilon'=\left\{\begin{array}{cc}
 1& \sigma\mbox{spin up}\\ -1& \sigma 
  \mbox{spin down}
  \end{array} \right. 
\end{array}
\end{equation}
This means that the hamiltonian (55) gives the coupled XY 
model discussed in the last section with the twisted boundary
\begin{equation}
\begin{array}{rcl}
\sigma^{\pm}_{L+1}&=&\displaystyle
     e^{-\epsilon(a_{\sigma}+c_{\sigma})}
     \exp\{\pm\epsilon[a_0+c_0+(a_{\sigma}+c_{\sigma})
     N_{\sigma}+(a_{\tau}+c_{\tau})N_{\tau}]\}
      \sigma^{\pm}_1\\[3mm]
\tau^{\pm}_{L+1}&=&\displaystyle
     e^{-\epsilon'(a_{\tau}-c_{\tau})}
     \exp\{\pm\epsilon[a_0-c_0+(a_{\sigma}-c_{\sigma})
     N_{\sigma}+(a_{\tau}-c_{\tau})N_{\tau}]\}
      \tau^{\pm}_1
\end{array}
\end{equation}
Comparing the Bethe Ansatz equations (37), (38), (53) and 
(54), we find that they are same if the free parameters are 
fixed 
\begin{equation}
a_{\sigma}=a_{\tau}=c_{\sigma}=-c_{\tau}=-\frac{i\pi}2,
a_0=iL\pi,					 c_0=i\pi \label{condition}
\end{equation}

Let us prove  the connection (58) by using the Jordan-Wigner 
transformation on transfer matrix. The Jordan-Wigner 
transformation on operators is defined by 
\begin{equation}
\left(\begin{array}{c}\sigma^+_{m}\\
    \sigma^-_{m} \end{array}\right)
= V^2_{m\uparrow}
\left(\begin{array}{c}a^+_{m\uparrow}\\a_{m\uparrow} \end{array}
  \right) 
=\left(\begin{array}{cc}v^2_{m\uparrow}&0\\0&v^{-2}_{m\uparrow} 
\end{array}\right) 
\left(\begin{array}{c}a^+_{m\uparrow}\\a_{m\uparrow} \end{array}
  \right) 
\end{equation}
 
\begin{equation}
\begin{array}{rcl}
\left(\begin{array}{c}\tau^+_{m}\\
    \tau^-_{m\uparrow} \end{array}\right)& =&
\displaystyle V^2_{m\downarrow}
\left(\begin{array}{c}a^+_{m\downarrow}\\a_{m\downarrow} \end{array}
  \right)\\[5mm]
&=& \displaystyle 
\left(\begin{array}{cc}v^2_{m\uparrow}u^2_{m\uparrow}r^2_m 
v^2_{m\downarrow}&0\\0&(v_{m\uparrow}u_{m\uparrow}r_m 
v_{m\downarrow})^{-2} \end{array}\right) 
\left(\begin{array}{c}a^+_{m\downarrow}\\a_{m\downarrow} \end{array}
  \right)
\end{array}
\end{equation}
with the definition
\begin{equation}
\begin{array}{rcl}
v_{ms}&=&\displaystyle \exp \{i\frac{\pi}2\sum_{k=1}^{m-1}
(n_{ks}-1)\} \\[3mm]
u_{ms}&=&\displaystyle \exp \{i\frac{\pi}2
(n_{ks}-1)\} \\[3mm]
r_{m}&=&\displaystyle \exp \{i\frac{\pi}2\sum_{k=m+1}^{L}
(n_{k\uparrow}-1)\} 
\end{array}
\end{equation}
Under the transformation,
the L-operator changes into \cite{OWA}:
\begin{equation}
\cl_m(\mu)=V_{m+1}L_m(\mu)V_m \label{loperator}
\end{equation}
where
\begin{equation}
\begin{array}{rcl}
V_{m+1}&=&V_m(U_{m\uparrow}\otimes U_{m\uparrow})\\
       &=&   V_{m\uparrow} U_{m\uparrow}\otimes 
          V_{m\downarrow}U_{m\downarrow}\\[5mm]
U_{m,s}&=&\mbox{dia}(u_{m,s}, u^{-1}_{m,s})
\end{array}
\end{equation}
Substituting equation (\ref{loperator}) into $t_H(\mu)$, 
we obtain
\begin{eqnarray}
t_H(\mu)&=&\ct_{11}(\mu)-\ct_{22}(\mu)-\ct_{33}(\mu)
           +\ct_{44}(\mu) \nonumber\\[5mm]
        &=&e^{\beta_1}T_{11}(\mu)+e^{\beta_2}T_{22}(\mu)
           +e^{\beta_3}T_{33}(\mu)
           +e^{\beta_4}T_{44}(\mu)
\end{eqnarray}
where
\begin{eqnarray}
e^{\beta_1}&=&\exp\{-i\frac{\pi}4\sum_{j=1}^L
              (\sigma_j^z+\tau_j^z-2)\}=e^{-\beta_4}\nonumber\\
e^{\beta_2}&=&\exp\{-i\frac{\pi}4\sum_{j=1}^L
              (\sigma_j^z-\tau_j^z)\}=e^{-\beta_3}
\end{eqnarray}
This is exactly equal to equation (46) with condition 
(\ref{condition}). This finished our proof. 

It is worthy to point out that Wadati et al \cite{OWA} also
applied the Jordan-Wigner transformation to the L-operator 
and the R matrix. In the derivation of the hamiltonian of 
the Hubbard model, they impose the periodic boundary
condition. But, they did not consider the relation
between the boundaries. 

In the rest of this section, we give an another independent 
proof in  terms of the hamiltonians. The periodic 
Hubbard model is
\begin{equation}
\begin{array}{rcl}
H&=&\displaystyle
   -\sum_{m=1,s}^{L-1}(a^+_{m+1,s}a_{m,s}+a^+_{m,s}a_{m+1,s})
  +\sum_{m=1,s}^{L}(n_{m\uparrow}-\frac12)(n_{m\downarrow}-\frac12)
  \\[3mm]
  & &\displaystyle -\sum_{s=\uparrow,\downarrow}
  (a^+_{1,s}a_{L,s}+a^+_{L,s}a_{1,s})
\end{array}
\end{equation}
where, we have use the periodic condition
\begin{equation}
a^+_{L+1,s}=a^+_{1,s}, a_{L+1,s}=a_{1,s}.
\end{equation}
Using the Jordan-Wigner 
transformation, we can obtain \begin{equation}
\begin{array}{rcl}
a^+_{m+1,\uparrow}a_{m,\uparrow}+a^+_{m,\uparrow}a_{m+1,\uparrow}
&=&-(\sigma^+_{m+1}\sigma^-_m+\sigma^+_{m}\sigma^-_{m+1})\\[3mm]
a^+_{m+1,\downarrow}a_{m,\downarrow}+a^+_{m,\downarrow}a_{m+1,
   \downarrow}
&=&\displaystyle 
-(\tau^+_{m+1}\tau^-_m+\tau^+_{m}\tau^-_{m+1})\\[3mm]
a^+_{1,\uparrow}a_{L,\uparrow}+a^+_{L,\uparrow}a_{1,\uparrow}
&=&\displaystyle \exp\{\frac{i\pi}2\sum_{j=1}^L(\sigma_j^z-\frac12)\}
 (\sigma^-_L\sigma^+_{1}+\sigma^+_{L}\sigma^-_{1})\\[3mm]
a^+_{1,\downarrow}a_{L,\downarrow}+a^+_{L,\downarrow}a_{1,
\downarrow}
&=&\displaystyle \exp\{\frac{i\pi}2\sum_{j=1}^L(\tau_j^z-\frac12)\}
(\tau^-_L\tau^+_1+\tau^+_{L}\tau^-_{1})
\end{array}
\end{equation}
The hamiltonian changes into
\begin{equation}
\begin{array}{rcl}
H&=&\displaystyle\sum_{m=1}^{L-1}(\sigma^+_{m+1}\sigma^-_{m}
    +\sigma^-_{m+1}\sigma^+_{m})-\exp\{\frac{i\pi}2\sum_{j=1}^L
    (\sigma_j^z-\frac12)\}(\sigma^-_L\sigma^+_{1}
    +\sigma^+_{L}\sigma^-_{1})\\[3mm]
&&\displaystyle+\sum_{m=1}^{L-1}(\tau^+_{m+1}\tau^-_{m}
    +\tau^-_{m+1}\tau^+_{m})-
     \exp\{\frac{i\pi}2\sum_{j=1}^L(\tau_j^z-\frac12)\}
(\tau^-_L\tau^+_{1}+\tau^+_{L}\tau^-_{1})\\[3mm]
& 
&\displaystyle + 
\frac{U}4\sum_{m=1}^{L}\sigma^z_m\tau^z_m
\end{array}
\end{equation}
Therefore, under the Jordan-Wigner transformation, 
the hamiltonian of the Hubbard model becomes one 
of the coupled XY model with twisted boundary condition
\begin{equation}
\begin{array}{rcl}
\sigma^{\pm}_{L+1}&=&\displaystyle
     \exp\{\pm\frac{i\pi}2\sum_{j=1}^L(\sigma^z_j-1)\}
      \sigma^{\pm}_1\\[3mm]
\tau^{\pm}_{L+1}&=&\displaystyle
     \exp\{\pm\frac{i\pi}2\sum_{j=1}^L(\tau^z_j-1)\}
     \tau^{\pm}_1
\end{array}
\end{equation}
which is coincident with equations (57) and (58).

\section{ Concluding remarks}

In this paper, we have found the  eigenvalues 
of the Hubbard model and the coupled XY model with 
twisted boundary condition by using the analytic 
Bethe Ansatz method. We have shown how they are equal 
on the levels hamiltonian and the transfer matrix by 
using the Jordan-Wigner transformation. 
The power expansion of $\log (t^g(\mu))$ in terms of  
$\mu$ will give explicitly  the infinite number of 
conserved  quantities. In this
level we claim the coupled XY model with 
twisted boundary condition is integrable.

It is worthy to point out that we consider 
here only a set of special 
boundary conditions. It is not difficult 
to generalize to other
kinds of twisted boundary conditions. 
For the open boundary, one should
consider the solution of the reflection equations. 
This will be related to the surface 
critical behaviour of the system.

{Note: After finishing this paper, we was told by 
Prof. Wadati that Ramos and Martin  \cite{RM} 
found the eigenvalue of the Hubbard model by using the 
algebraic Bethe Ansatz method, which recovers partly 
our results from the different approach. In \cite{RM}, 
they also notice the effect of boundary 
condition, but they did not discuss it in detail.


\begin{thebibliography}{[99]}
\bibitem{LW} E. Lieb and F.Y. Wu, Phys. Rev. Lett. {\bf 20}  (1968)1445.
\bibitem{Sha1}B.S. Shastry, Phys. Rev. Lett. {\bf 56} (1986)1529.
\bibitem{Sha2}B.S. Shastry, Phys. Rev. Lett. {\bf 56} (1986)2453.
\bibitem{WOA}M. Wadati, E. Olmedilla and Y. Akutsu, 
    J. Phys. Soc. Jpn. {\bf 56}(1987)1340.
\bibitem{OWA} E. Olmedilla, M. Wadati and Y. Akutsu, 
    J. Phys. Soc. Jpn. {\bf 56}(1987)2298.
\bibitem{OW} E. Olmedilla and  M. Wadati,  
    Phys. Rev. Lett. {\bf 60} (1987)1595.
\bibitem{SW}M. Shiroishi and M. Wadati, J. Phys. Soc. Jpn. 
  {\bf 64}(1996)57.
\bibitem{Sha3}B.S. Shastry, J. Stat. Phys. {\bf 50} (1988)1340.
\bibitem{Res1}N.Yu. Reschetikhin, Sov. Phys. Jept. {\bf 57} 
(1983)691.
\bibitem{VR}V.I. Virchirko and N.Yu. Reschetikhin, Theor. Mat. Fiz
{\bf 56}(1983)260.
\bibitem{Res2}N.Yu. Reschetikhin, Phys. Lett. {\bf B56} (1991)133.
\bibitem{Mar} M.J. Martins, Phys. Rev. Lett. {\bf 74}(1995)3316,
Phys. Lett. {\bf B359}(1995)334.
\bibitem{Bar}R.Z. Bariev, Theor. Mat. Fiz. {\bf 82}(1990)313.
\bibitem{RM}P.B. Ramos and M.J. Martins, Preprint (1996) 
UFSCARF-TH-96-10.

\end{thebibliography}
\end{document}